\RequirePackage{fix-cm}
\RequirePackage{amsmath}
\documentclass[twocolumn,epjc3]{svjour3}  
\smartqed  
\RequirePackage{graphicx}
\RequirePackage{mathptmx}      
 
\usepackage{color}
\usepackage{mathrsfs}
\usepackage{amssymb, bm}

\newcommand{\be}{\begin{equation}}
\newcommand{\ee}{\end{equation}}

\begin{document}

\title{When can we compute analytically lookback time, age of the 
universe, and luminosity distance?}

\author{Sonia Jose\thanksref{e1,addr1}
	   \and
Alexandre Leblanc  \thanksref{e2,addr2} 
	   \and
Valerio Faraoni  \thanksref{e3,addr1}}


\thankstext{e1}{e-mail: sjose21@ubishops.ca}
\thankstext{e2}{e-mail: alexandre.leblanc3@usherbrooke.ca}
\thankstext{e3}{e-mail: vfaraoni@ubishops.ca}


\institute{Department of Physics \& Astronomy, Bishop's University, 
2600 College Street, Sherbrooke, Qu\'ebec, Canada J1M~1Z7 \label{addr1}
           \and
Department of Physics, Universit\'e de Sherbrooke, 2500 
Boulevard de l'Universit\'e, Sherbrooke, Qu\'ebec, Canada J1K 2R1  
\label{addr2}
}

\date{Received: date / Accepted: date}

\maketitle

\begin{abstract} 

In Friedmann--Lema\^itre--Robertson--Walker cosmology, it is sometimes 
possible to compute analytically lookback time, age of the universe, and 
luminosity distance versus redshift, expressing them in terms of a finite 
number of elementary functions. We classify these situations using the 
Chebyshev theorem of integration and providing examples.


\end{abstract}

\section{Introduction}
\label{sec:1}
\setcounter{equation}{0}

When can we compute {\em exactly} lookback time, age of the universe, and 
redshift-luminosity distance relation $D_L(z)$ in 
Friedmann--Lema\^itre--Robertson--Walker (FLRW)~ cosmology? These 
quantities, of crucial importance for modern cosmology, are expressed by 
integrals taking the form of infinite hypergeometric series.  Of course, 
lookback time, age, and luminosity distance can always be computed 
numerically in a given cosmological model, however one would also like to 
know when they can be computed analytically in terms of a finite number of 
elementary functions. This simplification happens when the hypergeometric 
series expressing them truncate. Equivalently, it happens when the 
integral expressing lookback time, age, or luminosity distance is of a 
special form contemplated by the Chebyshev theorem of integration 
\cite{Chebyshev,MarchisottoZakeri}. The truncation of the hypergeometric 
series, or the equivalent Chebyshev theorem, were used in the 1960s 
\cite{Jacobs1968,Vajik69,McIntosh1972,McIntoshFoyster1972}, and were 
recently rediscovered \cite{Chen:2014fqa}, to derive two- and three- fluid 
(or effective fluid) analytical solutions of the Einstein-Friedmann 
equations (see \cite{Faraoni:2021opj} for a review).

When the matter content of the FLRW universe consists of at most three 
non-interacting fluids or effective fluids (which includes spatial 
curvature and/or the cosmological constant $\Lambda$, if present), and 
assuming that 
the equations of state of these fluids are of the form $P=w\rho$ with $w$ 
constant and rational, the situations in which lookback time, age, and 
luminosity distance are analytical and simple are classified by means of 
the Chebyshev theorem \cite{Chebyshev,MarchisottoZakeri}.

The assumption that the equation of state parameter $w$ be a rational 
number is not restrictive. First, this is almost always a rational number 
in the cosmological literature 
\cite{Waldbook,EllisMaartensMacCallum,KolbTurner,Slava}. Second, 
even if $w$ is irrational, in practice cosmological observations cannot 
distinguish between $w$ and its rational approximation and it is an 
excellent approximation to replace the actual value of this parameter with 
a rational approximation to it containing a sufficient number of digits.

Sections~\ref{sec:2} and~\ref{sec:3} catalogue situations in which the 
universe is characterized by two or three fluids or effective fluids 
(which includes cosmological constant and spatial curvature) and lookback 
time and age are computed analytically in simple form. Section~\ref{sec:4} 
reports situations in which one can compute analytically the luminosity 
distance $D_\mathrm{L}(z)$ as a function of redshift $z$. 
Section~\ref{sec:5} contains concluding remarks.

We follow the notation and conventions of Ref.~\cite{Waldbook}: the metric 
signature is ${-}{+}{+}{+}$, $G$ is Newton's constant, and units are used 
in which the speed of light $c$ is unity. 

\section{Lookback time and age of the universe}
\label{sec:2}

We consider a homogeneous and isotropic universe described by the FLRW 
line element in spherical comoving coordinates $\left(t,r, 
\vartheta,\varphi \right)$
\begin{equation}
 ds^2=-dt^2 +a^2(t)\left( \frac{dr^2}{1-Kr^2} + r^2 d\Omega_{(2)}^2 
\right)
\end{equation}
where $d\Omega_{(2)}^2=d\vartheta^2 +\sin^2  \vartheta \, d\varphi^2$ is 
the line element on the unit 2-sphere, $K$ is the curvature index (which 
we take to be normalized to $0, \pm 1$), and $a(t)$ is the scale factor. 
We assume that the
matter source of the FLRW universe is a perfect 
fluid with energy density $\rho$ and pressure $P$ related by the 
barotropic, linear, and constant equation of state 
\be
P=w\rho \,, \quad\quad  w=\mbox{const.}
\ee
The Einstein-Friedmann equations describing the evolution of this universe  
read 
\begin{eqnarray}
&& H^2= \frac{8\pi G}{3} \, \rho +\frac{\Lambda}{3}-\frac{K}{a^2} 
\,,\label{Friedmann}\\
&&\nonumber\\
&& \frac{\ddot{a}}{a} = -\frac{4\pi G}{3} \left( \rho +3P \right) 
+\frac{\Lambda}{3} \,,\label{acceleration}\\
&&\nonumber\\
&& \dot{\rho}+3H\left( P+\rho \right) =0 \,, \label{conservation}
\end{eqnarray}
where an overdot denotes differentiation with respect to the comoving 
time $t$, $H \equiv \dot{a}/a$ is the Hubble function and $\Lambda$ is the 
cosmological constant.   The conservation equation~(\ref{conservation}) 
gives 
\be
\rho(a)= \rho_0 \left( \frac{a_0}{a} \right)^{3(w+1)}  \,, 
\label{rhoscaling}
\ee
where $\rho_{0}$ is a constant. Suppose that the cosmic fluid is  a 
mixture of $n$ non-interacting 
fluids with individual densities $\rho_{i}$ and 
pressures $P_i$,  with $ P_i = w_i \rho_i $ and $ 
w_i=$~const. ($i=1,2,3, \, ... \, , n$). The total energy density and 
pressure are 
\be
\rho_\mathrm{(tot)}=\sum_{i=1}^n \rho_i \,, \quad\quad 
P_\mathrm{(tot)}= \sum_{i=1}^n w_i \rho_i \,,
\ee
respectively. 

We consider universes beginning with a Big Bang $a(0)=0$ at $t=0$ and 
we denote the present value of time-dependent quantities with a zero 
subscript. Then, since $\dot{a} = da/dt$, the lookback time to a source 
that emitted at time $t_e$, scale factor $a_e$, and redshift $z_e$ is  
\be
t_L = \int_{t_e}^t dt'= \int_{a_e}^{a_0}  \frac{da'}{ \dot{a}' }
\ee
with $ \dot{a} $  given by the Friedmann 
equation~(\ref{Friedmann}).  In the limit $t_e \to 0$ (or $a_e\to 0$, or 
$z_e\to+\infty$), one obtains the age of the universe
\be
t_0 = \int_0^t dt'= \int_0^{a_0}  \frac{da'}{ \dot{a}' } \,.
\ee
Rewrite the Friedmann equation as  
\be
\frac{K}{a^2} =H^2 \left( \Omega^\mathrm{(tot)} +\Omega_{\Lambda}  -1 
\right) \,,
\ee
where $\Omega^\mathrm{(tot)}$ is the total energy density of the 
real fluids (as opposed to the {\em effective} fluids given by $\Lambda$ 
and by the curvature term) in units of the 
critical 
density $\rho_c \equiv \frac{3H^2}{8\pi G}$. For a single fluid,  
using Eq.~(\ref{rhoscaling}) one obtains
\begin{eqnarray}
\dot{a} &=& \sqrt{ \frac{8\pi G}{3} \, a_0^2 \rho_0 \left( 
\frac{a_0}{a}\right)^{3w+1} - \left( \Omega_0 -1\right) a_0^2 H_0^2 
+\frac{\Lambda a_0^2}{3} \left( \frac{a_0}{a}\right)^{-2} } \nonumber\\
&&\nonumber\\
& = & a_0 H_0 \sqrt{ 1-\Omega_0^\mathrm{(tot)}  + \Omega_0 
\left( z+1 \right)^{3w+1} + \Omega_{\Lambda0} \left( z+1 \right)^{-2} } 
\,, \nonumber\\
&&
\end{eqnarray}
where $z \equiv a_0/a -1 $ is the redshift factor, and then  
\begin{eqnarray}
t_L &=& H_0^{-1} \int_0^{z_e} dz' \left( z'+1\right)^{-2} \left[ 
1-\Omega_0^\mathrm{(tot)}+\Omega_0 \left( z'+1\right)^{3w+1} 
\right.\nonumber\\
&&\nonumber\\
&\, & \left.+\Omega_{\Lambda 0}\left( z'+1\right)^{-2} \right]^{-1/2} \,.
\end{eqnarray}
The change of variable $ z \to  x \equiv a/a_0 = \left( 1+z \right)^{-1} $ 
in the integral turns it into
\be
t_L =H_0^{-1} \int_{x_e}^1   
\frac{dx}{\sqrt{1-\Omega_0^\mathrm{(tot)} + \Omega_0 \, 
x^{ -(3w+1)} + \Omega_{\Lambda 0} \, x^2 } } \,. \label{general}
\ee
For suitable values of the equation of state parameter $w$, this integral 
can be expressed in terms of elementary functions using the Chebyshev 
theorem of  integration \cite{Chebyshev,MarchisottoZakeri}:\\\\ 
{\em The integral
\be
J \equiv \int dx \, x^p \Big( \alpha + \beta x^r\Big)^q \,,
\quad\quad \quad 
r\neq 0 \,, \quad p,q,r \in \mathbb{Q} \label{integral}
\ee
admits a representation in terms of elementary functions if and only if 
at least one of ~$ \frac{p+1}{r}$, ~$q$, ~$\frac{p+1}{r}+q$ is an 
integer.} 

In order for the integral in Eq.~(\ref{general}) to be of the Chebyshev 
form, one of the following possibilities needs to be realized.

\subsection {$K = 0$, $\Lambda \neq 0$, plus a single fluid}

Suppose that the universe is sourced by a single fluid with equation of 
state parameter $w$ and has non-zero cosmological constant $\Lambda$. This 
situation includes the $\Lambda$-Cold Dark Matter ($\Lambda$CDM) model if 
the fluid is a dust. Spatial flatness $K=0$ is equivalent to 
$\Omega_0^\mathrm{(tot)} =1 $  and the integral in 
Eq.~(\ref{general}) has the form\footnote{If $\Lambda=0$ the integration 
is trivial.}
\be
t_L \, H_0 = \int_{x_e}^1 dx \, x^{-1} \left[ \Omega_{0} \, x^{-3(1+w)} 
+\Omega_{\Lambda 0} \right]^{-1/2}   \label{strakaz}
\ee
{\em i.e}, the form~({\ref{integral}}) with
\be
p=-1\,, \quad\quad  r=-3(1+w)\neq 0\,, \quad\quad 
q=-1/2 \,,
\ee
which all are rational if $w$ is. Most walues of $w$ considered in the 
cosmological literature are rational but, in any case, cosmological 
observations cannot distinguish between an irrational value of $w$ and its 
rational approximation, hence in practice one can always assume $w\in 
\mathbb{Q}$.  The Chebyshev theorem applies since $\frac{p+1}{r} = 0$. 
Indeed, a direct computation of the integral gives the lookback time 
(see~\ref{sec:appendix})
\begin{eqnarray}
t_L &=& \frac{H_0^{-1} }{3(w+1) \sqrt{\Omega_{\Lambda 0} } } \left[ 
\ln \Bigg( \frac{ 1+\sqrt{\Omega_{\Lambda 0} } }{1-\sqrt{\Omega_{\Lambda 
0}} } \Bigg) \right. \nonumber\\
&&\nonumber\\
&\, & \left. 
-\ln \Bigg( 
\frac{ \sqrt{\Omega_{\Lambda 0}\, x_e^{3(w+1)} +\Omega_0}      
+  \sqrt{\Omega_{\Lambda 0}\, x_e^{3(w+1)} }     }{
\sqrt{\Omega_{\Lambda 0}\, x_e^{3(w+1)} +\Omega_0} -
\sqrt{\Omega_{\Lambda 0}\, x_e^{3(w+1)} } } \Bigg) \right] \,.
\end{eqnarray}
In the limit $x_e\to 0$, $t_L\to 0$ and we obtain the age of the universe 
(see~\ref{sec:appendix})
\be
t_0 =\frac{2H_0^{-1}}{3(w+1) \sqrt{\Omega_{\Lambda 0}}} \, \ln \Bigg( 
\frac{ 1+ \sqrt{\Omega_{\Lambda 0}} }{\sqrt{ 1- \Omega_{\Lambda 0}} } 
\Bigg) \,.
\ee
This formula appears in textbooks ({\em e.g.}, 
\cite{KolbTurner,EllisMaartensMacCallum}) for the 
special case $w=0$ of dust.

\subsection{$\Lambda=0$, single fluid plus spatial curvature}

This situation also leads to physically interesting scenarios. Some of 
these universes contain dust or radiation and are found in cosmology 
textbooks.

For a single fluid with equation of state parameter $w$ and spatial 
curvature ($K = \pm 1$), the lookback time~(\ref{general}) is 
\be
t_0 = H_0^{-1} \int_{x_e}^1 dx \,  
\left[ 1-\Omega_0^\mathrm{(tot)} + \Omega_0 \, x^{-\left( 3w+1 \right)} 
\right]^{-1/2} 
\,.
\ee
Comparing with Eq. ~(\ref{integral}) yields the exponents 
\be
p=0 \,, \quad\quad r=-(3w+1) \,, \quad\quad q=-1/2 \,, 
\ee
which are all rational if $w \in \mathbb{Q}$. The  conditions for the  
Chebyshev theorem to hold are\\
\be
\frac{p+1}{r}=\frac{-1}{3w+1} = n   
\ee
or
\be
\frac{p+1}{r}+q=\frac{-3(w+1)}{2(3w +1)} = m \,,
\ee
where $n, m \in \mathbb{Z}$. The possible values of $w$ for this to 
happen are the countable infinities of values 
\be
w_n =- \frac{(n+1)}{3n} \label{omega}
\ee
and 
\be
w_m =-\frac{(3+2m)}{3(1+2m)} \,. \label{wm}
\ee
Only a few of the equation of state parameters thus obtained are 
of physical interest. Focusing on the first possibility~(\ref{omega}), as 
$n$ spans the values 
$ n=-\infty , \, ... \, , -3, -2, -1, 1, 2, 3, \, ... \,, +\infty $, 
one obtains
\be
w_n =-\frac{1}{3}, \, ... \,,  -\frac{2}{9}, -\frac{1}{6},\;\;0, - \frac{ 
2}{3}, -\frac{1}{2}, - \frac{4}{9},  \, ... \,, -\frac{1}{3}\,.
\ee
By imposing the second condition~(\ref{wm}), as $m$ spans the range  
$ -\infty , \, ... \,, -3, 
-2, -1, 1, 2, 3, \, ... \,, +\infty$, one obtains 

\be
w_m=-\frac{1}{3}, \, ... \,,  -\frac{1}{5}, 
-\frac{1}{9},\;\;\frac{1}{3},\;\; -1,  - \frac{5}{9}, 
-\frac{7}{15},   \, ... \,, -\frac{1}{3} \,.
\ee
The values $w=0$ (dust), $w=1/3$ (radiation), and $w=-1/3$ (empty space 
with a hyperbolic foliation, {\em i.e.}, the Milne universe) correspond to 
textbook cases 
\cite{EllisMaartensMacCallum,Waldbook,KolbTurner,Slava}. For dust, they 
give the well-known Mattig relation\footnote{This relation can also be 
derived from the geodesic deviation equation for 
null geodesics \cite{Ellis:1998ct}.} 
\cite{Mattig58}
\begin{eqnarray}
D_L(z) &=& \frac{2H_0^{-1} }{\Omega_0^\mathrm{(dust)} } \nonumber\\
&&\nonumber\\
& \times & \left[
\Omega_0^\mathrm{(dust)} \, z - \left(2- \Omega_0^\mathrm{(dust)} \right) 
\left( \sqrt{1+\Omega_0^\mathrm{(dust)}\, z} -1 \right)\right] 
\,.\nonumber\\
&&
\end{eqnarray}

The values of $w$ different from $0, \pm 1/3$ found above describe phantom 
or quintessence fluids that, although unrealistic to describe the present 
universe, could be used as toy models for theoretical purposes. The 
degenerate case $w =-1$ reproduces the empty universe with cosmological 
constant and spatial curvature.

\section{Lookback time and age of a spatially flat universe with two (real 
or effective) fluids}
\label{sec:3}

Consider the case of two fluids with equations of state $P_1=w_1\rho_1$ 
and $P_2=w_2\rho_2$, with $w_{1,2} =$ constants. It is assumed that these 
two fluids have the same four-velocity $u^{c}$ in their stress-energy 
tensors. We regard cosmological constant and spatial curvature term as 
effective fluids hence, in the following, certain conditions correspond to 
the possibility of one or both fluids being the curvature- or the 
$\Lambda$- (effective) fluids. The total fluid density is 
$\rho_\mathrm{(tot)}=\rho_{1}+\rho_{2}$ and the individual densities scale 
as
\be
\rho_1= \rho_1^{(0)}  \left( \frac{a_0}{a} \right)^{3(w_1+1)} \,, 
\quad\quad 
\rho_2= \rho_2^{(0)}  \left( \frac{a_0}{a} \right)^{3(w_2+1)} \,,
\ee
then the integral in Eq.~(\ref{general}) is
\begin{align}
   t_0 H_0 = \int_0^1 dx \, x^{(3w_1+1)/2} \left[ 
   \Omega_{0}^{(1)} + \Omega_{0}^{(2)} \, x^{3(w_1 - w_2)}  \right]^{-1/2}
\end{align}
and comparison with Eq. ~(\ref{integral}) yields the exponents 
\be
p=\frac{3w_1 + 1}{2} \,, \quad\quad r=3(w_1 - w_2) \,, \quad\quad q=-1/2 
\,, 
\ee
with $w_1\neq w_2$. The conditions for the Chebyshev theorem to hold are\\ 
\be
\frac{p+1}{r}=\frac{w_1+1}{2(w_1-w_2)} = n  \in \mathbb{Z}  \label{racci1}
\ee
or
\be
\frac{p+1}{r}+q=\frac{w_2+1}{2(w_1-w_2)} = m \in \mathbb{Z} \,. 
\label{seccondition}
\ee

\subsection{First condition: $  \frac{p+1}{r}=n \in \mathbb{Z} $}

The first integrability condition~(\ref{racci1}) gives
\be
\left(2n-1\right) w_1 -2nw_2 =1 \quad\quad \mbox{if} \:\: w_1\neq w_2 
\label{acci1}
\ee
(the case $w_1=w_2$ corresponds to $n=\pm \infty$).  
Fixing the first fluid ({\em i.e.}, $w_1\in \mathbb{Q}$) yields integrable  
cases by varying $n$.

\subsubsection{Dust plus a second (real or effective) fluid}

If the first fluid is a dust with  $w_1=0$, ``simple'' integrability 
cases are obtained when 
\be
w_2=-\frac{1}{2n} \,;
\ee
as $n=-\infty, \, ... \, , -3, -2, -1, 1, 2, 3, \, ... \,, +\infty$, 
we obtain the  pairs
\be
\begin{array}{rcll}
\left( w_1, w_2 \right) &=& \left( 0, 0 \right) & \mbox{(single dust 
fluid)},\\
&&&\\
 & ... &   & \\
&&&\\
\left( w_1, w_2 \right) &=& \left( 0,  \pm \frac{1}{6} \right) ,& \\
&&&\\
\left( w_1, w_2 \right) &=& \left( 0, \pm \frac{1}{4}  \right) ,&\\
&&&\\
\left( w_1, w_2 \right) &=& \left( 0, \pm \frac{1}{2} \right) ,&\\
&&&\\
&  ... & &\\
&&&\\
\left( w_1, w_2 \right) &=& \left( 0 , 0\right) & \mbox{(again, a 
single dust fluid)}.
\end{array}\label{list1}
\ee
Most of the analytical cases shown above do not have much 
physical relevance. Although quintessence models are present, the values 
of $w_2$ are not close to the value $-1$ measured by current observations. 
The limit $n\rightarrow  \pm \infty$ produces a second dust, {\em i.e.}, 
there is a single dust fluid in the FLRW universe and integration is 
trivial.

\subsubsection{Radiation plus a second (real or effective) fluid}

If the first fluid is radiation, $w_1=1/3$, the corresponding values of 
$w_2$ for integrability \`a la Chebyshev are
\be
w_2=\frac{n-2}{3n} \,,
\ee
producing the pairs
\be
\begin{array}{rcll}
\left( w_1, w_2\right) &=&  \left( \frac{1}{3}  , \frac{1}{3} \right) 
&\mbox{(a single radiation fluid)}, \\
&&&\\
&  ... & &\\  
&&&\\
\left( w_1, w_2 \right) &=&  \left( \frac{1}{3}  , \frac{5}{9} 
\right) ,&\\
&&&\\ 
\left( w_1, w_2 \right) &=&  \left( \frac{1}{3} , \frac{2}{3} \right) ,&\\
&&&\\ 
\left( w_1, w_2 \right) &=&  \left( \frac{1}{3}  ,  1 \right) 
&\mbox{(radiation plus stiff fluid)}, \\
&&&\\ 
\left( w_1, w_2 \right) &=&  \left( \frac{1}{3}  , -\frac{1}{3} \right) & 
\mbox{(radiation plus spatial curvature)},\\ 
&&&\\ 
\left( w_1, w_2 \right) &=& \left( \frac{1}{3}  ,  0 \right) & 
\mbox{(radiation plus dust)},\\
&&&\\ 
\left( w_1, w_2 \right) &=&  \left( \frac{1}{3}  ,  \frac{1}{9} \right) ,&\\
&&&\\
&  ...& &\\
&&&\\ 
\left( w_1, w_2 \right) &=&   \left( \frac{1}{3}  , \frac{1}{3} \right) 
&\mbox{(again, a single  radiation fluid)}.
\end{array}\label{list2}
\ee

\subsubsection{Cosmological constant plus a second (real or effective) 
fluid}

The value $w_1=-1$ corresponds to $n=0$ in Eq.~(\ref{racci1}) and is 
satisfied by any value of $w_2 \neq -1$, producing a cosmological 
constant with any single perfect fluid.

\subsubsection{Stiff matter plus a second (real or effective) fluid}

In this case $w_1=1$, 
\be
w_2= \frac{n-1}{n} \,,
\ee
and we have the pairs 
\be
\begin{array}{rcll}
\left( w_1, w_2 \right) &=& \left( 1, 1 \right) & 
\;\;\;\;\mbox{(single stiff fluid)},\\
&&&\\
&  ... &   &\\
&&&\\
\left( w_1, w_2 \right) &=& \left( 1, \frac{4}{3}  \right) ,& \\
&&&\\
\left( w_1, w_2 \right) &=& \left( 1, \frac{3}{2}  \right) ,&\\
&&&\\
\left( w_1, w_2 \right) &=& \left( 1,  2\right) ,&\\
&&&\\
\left( w_1, w_2 \right) &=& \left( 1, 0 \right) &  \mbox{(stiff 
matter plus dust)},\\
&&&\\
\left( w_1, w_2 \right) &=& \left( 1, \frac{1}{2} \right) ,&\\
&&&\\
\left( w_1, w_2 \right) &=& \left(  1, \frac{2}{3} \right) ,&\\
&&&\\
&  ... & \\
&&&\\
\left( w_1, w_2 \right) &=& \left( 1 , 1 \right) &\mbox{(again, a 
single stiff fluid)}.
\end{array} \label{list3}
\ee
The physically most relevant situation is that of a  stiff fluid plus 
dust.

\subsection{Second condition: $ \frac{p+1}{r} +q =m \in \mathbb{Z} $}

The second condition (\ref{seccondition})  yields
\be
w_2=\frac{2m w_1 -1}{2m+1}  \quad \mbox{if} \: w_2 \neq w_1 \,;  
\label{acci2}
\ee
as done for the first condition, we fix the first fluid ({\em 
i.e.}, $w_1\in \mathbb{Q}$), and we obtain integrable cases as $m$ 
varies.

\subsubsection{Dust plus a second (real or effective) fluid}

If the first fluid is a dust, $w_1=0$, we have 
\be
w_2=-\frac{1}{1+2m} 
\ee 
and the pairs
\be
\begin{array}{rcll}
\left(w_1, w_2\right) &=&   \left(0 ,0 \right) & \mbox{(a 
single dust fluid)},\\
&&&\\
& ... & & \\  
&&&\\
\left(w_1, w_2\right) &=&   \left( 0 , \pm \frac{1}{5} \right) ,\\ 
&&&\\
\left(w_1, w_2\right) &=&    \left( 0 , \pm \frac{1}{3} \right) & 
\mbox{(dust plus radiation or curvature)}, \\
&&&\\
\left(w_1, w_2\right) &=&   \left( 0 , \pm 1 \right) & (\mbox{dust plus 
stiff fluid or} \; \Lambda ),\\
&&&\\
& ... & &\\
&&&\\
\left(w_1, w_2\right) &=&   \left( 0 , 0 \right)  & \mbox{(again, 
a single dust fluid)}. 
\end{array} \label{list4}
\ee
Physically plausible combinations include dust and 
stiff fluid, dust and radiation, dust and spatial curvature.

\subsubsection{Radiation plus a second (real or effective) fluid}

Beginning with radiation $w_1=1/3$, we obtain 
\be
w_2=\frac{2m-3}{3(1+2m)}
\ee
and the pairs
\be
\begin{array}{rcll}
\left(w_1, w_2\right) &=&  \left( \frac{1}{3}  , \frac{1}{3} \right) 
& \mbox{(single radiation fluid )},\\
&&&\\
& ... & &\\
&&&\\
\left(w_1, w_2\right) &=&   \left( \frac{1}{3} , \frac{3}{5} \right) ,&\\ 
&&&\\
\left(w_1, w_2\right) &=&  \left( \frac{1}{3}  , \frac{7}{9} \right) ,&\\
&&&\\
\left(w_1, w_2\right) &=&   \left( \frac{1}{3} , \frac{5}{3} \right) ,&\\
&&&\\
\left(w_1, w_2\right) &=&   \left( \frac{1}{3}  , 1 \right) &  
\mbox{(radiation plus stiff fluid)},\\ 
&&&\\
\left(w_1, w_2\right) &=&   \left( \frac{1}{3}  , -\frac{1}{9} \right) ,&\\
&&&\\
\left(w_1, w_2\right) &=&   \left( \frac{1}{3}  , \frac{1}{15} \right) 
,&\\
&&&\\
\left(w_1, w_2\right) &=&   \left( \frac{1}{3}  , \frac{1}{7} \right) ,&\\
&&&\\
& ... & &\\
&&&\\
\left(w_1, w_2\right) &=&    \left( \frac{1}{3}  , \frac{1}{3} \right)  
&\mbox{(again, a single radiation fluid)}. 
\end{array} \label{list5}
\ee

\subsubsection {$\Lambda$ plus a second (real or effective) 
fluid}

Setting $w_2=-1$ corresponds to $m=0$ and Eq.~(\ref{seccondition}) is 
satisfied for any $w_1\neq -1$.

\subsubsection{Stiff matter plus a second (real or effective) fluid}

Setting $w_1=1$ (the equation of state parameter of a stiff fluid) yields 
\be
w_2=\frac{2m-1}{2m+1}
\ee
and the pairs
\be
\begin{array}{rcll}
\left( w_1, w_2 \right) &=& \left( 1, 1 \right)& \mbox{(single stiff 
fluid)},\\
&&&\\
&  ... &  &\\
&&&\\
\left( w_1, w_2 \right) &=& \left( 1, \frac{7}{5}  \right) ,& \\
&&&\\
\left( w_1, w_2 \right) &=& \left( 1,  \frac{5}{3} \right) ,&\\
&&&\\
\left( w_1, w_2 \right) &=& \left( 1, 3 \right) ,&\\
&&&\\
\left( w_1, w_2 \right) &=& \left( 1, -1 \right) & \mbox{(stiff fluid plus  
$\Lambda$)},  \\
&&&\\
\left( w_1, w_2 \right) &=& \left( 1, \frac{1}{3} \right) &
\mbox{(stiff fluid plus radiation)},\\
&&&\\
\left( w_1, w_2 \right) &=& \left( 1, \frac{3}{5} \right) ,&\\
&&&\\
\left( n, w_2 \right) &=& \left(  1, \frac{5}{7} \right) ,&\\
&&&\\
&  ... & &\\
&&&\\
\left( w_1, w_2 \right) &=& \left( 1 , 1 \right) & \mbox{(again, a 
single stiff fluid)}.
\end{array}\label{list6}
\ee

\section{Luminosity distance}
\label{sec:4}

The luminosity distance versus redshift relation $D_L(z)$ is important to 
reconstruct the universe model from observations and has led to the 
discovery of the present acceleration of the cosmic expansion using type 
Ia supernovae 
\cite{SupernovaSearchTeam:1998fmf,SupernovaCosmologyProject:1998vns,Filippenko:1998tv,Riess:1999ti,Riess:2000yp,SupernovaSearchTeam:2001qse,SupernovaSearchTeam:2003cyd,SupernovaCosmologyProject:2003dcn,Barris:2003dq,SupernovaSearchTeam:2004lze,Riess:2019qba}. 
In addition, the reciprocity relation $D_L =\left(1+z\right)^2 D_A$ 
between luminosity distance $D_L $ and area distance $D_A$ is used as a 
probe of fundamental cosmology \cite{Bassett:2003vu}.

The luminosity distance in a FLRW universe is expressed by an integral of 
the Chebyshev form~(\ref{integral}) and can be calculated exactly in 
certain cases that we find below. First, let us review the derivation of 
$D_L(z)$ ({\em e.g.}, \cite{EllisMaartensMacCallum,Carroll}).

Let us rewrite the FLRW line element as
\be
ds^2 =- dt^2 +a^2(t) \left[ d\chi^2 +S_k^2(\chi) d\Omega_{(2)}^2 
\right] \,,
\ee
where $\chi$ is the hyperspherical radius and 
\be
S_k(\chi)= \left\{ \begin{array}{ccc}
\sin\chi & \quad \mbox{if} & K=1 \,,\\
&&\\
\chi & \quad \mbox{if} & K=0 \,,\\
&&\\
\sinh\chi & \quad \mbox{if} & K=-1 \,.\end{array} \right.
\ee
The luminosity distance $D_L$ between a light source and an observer is 
defined by 
\begin{align}
    D_L^2 = \frac{L}{4\pi F} \,,
\end{align}
where $L$ is the absolute luminosity of the source and $F$ is the flux 
density measured by the observer. Since
\be
\frac{F}{L} =\frac{1}{A\left( 1+z\right)^2} 
\ee
and the present area $A$ of a sphere of hyperspherical radius  
$\chi$ is $ A = 4\pi \, a_0^2 \, S_k^2(\chi) $, 
the luminosity distance becomes
\be
D_L = \sqrt{ \frac{A(1+z)^2}{4\pi}} 
= \left( 1+z \right) \, a_0  \, S_k (\chi) \,.
\ee
$\chi$ needs to be eliminated using
\be
\chi = a_0^{-1} \int \frac{da}{a^2 \, H(a)} =a_0^{-1} \int_0^z 
\frac{dz'}{H(z')} \,,
\ee
while the Einstein-Friedmann equation gives
\begin{eqnarray}
H^2 &=& \frac{8\pi G}{3} \, \sum_i \rho_i -\frac{K}{a^2} =  
\frac{8\pi G}{3} \, \sum_i \rho_i - a_0^2 H_0^2 \left( 
\Omega_0^\mathrm{(tot)} -1\right) \nonumber\\
&&
\end{eqnarray}
and $\rho_i = \rho_{i0} \left( 1+z \right)^{3(w_i+1)}$. Then, 
\be
H^2 = H_0^2 \sum_i \, \frac{\rho_i}{\rho_c} + \left( 
1-\Omega_0^\mathrm{(tot)} \right) a_0^2 \, H_0^2 \,,
\ee
where $\rho_c\equiv 3H^2 /(8\pi G)$ is the critical density. Finally,
\begin{eqnarray}
H(z) &=& H_0 \sqrt{ \sum_i \Omega_{i0} \left( 1+z \right)^{3(w_i+1)} 
+1-\Omega_0^\mathrm{(tot)} } \equiv H_0 \, E(z) \nonumber\\
&& 
\end{eqnarray}
gives $ 
\chi=  a_0^{-1} \int_0^z \frac{dz'}{E(z')} $
and the luminosity distance becomes  
\be
D_L(z) = (1+z) \, a_0 \, S_k \Bigg( 
\frac{1}{a_0 H_0} \int_0^z \frac{dz'}{E(z')} \Bigg) \,.
\ee
Since $ a_0= H_0^{-1}/\sqrt{ |\Omega_{K0}|}$,  
where $\Omega_{K0}= -\frac{K}{a_0^2 H_0^2} $ is the curvature density at 
the 
present time in units of the critical density, we have
\be
D_L (z) = \frac{(1+z) \, H_0^{-1} }{ \sqrt{ | \Omega_{K0}|} } \, S_K 
\Bigg( \sqrt{ 
|\Omega_{K0}| } \, \int_0^z \frac{dz'}{E(z')} \Bigg) \,,
\ee
which becomes particularly simple in a spatially flat universe 
\be
D_L^\mathrm{(flat)} (z) = (1+z) \, H_0^{-1} 
\,   \int_0^z \frac{dz'}{E(z')}  \,.\label{DLK=0}
\ee
Now the question is: can we express the integral
\be
I \equiv \int_0^z \frac{dz'}{E(z')} = \int_0^z \frac{dz'}{\sqrt{ 
1-\Omega_0^\mathrm{(tot)} + \sum_i \Omega_{i0}  (1+z')^{3(w_i+1)}  } } 
\ee
in terms of  a finite number of elementary functions?  This integral is 
similar to the one appearing in the lookback time~(\ref{general}), but now 
the limits of integration are $0$ and $z$ instead of $0$ and $1$.

\subsection{Single fluid}

For a single fluid, using the variable $x \equiv (1+z)^{-1}$, we have 
\begin{eqnarray}
I_1 & \equiv & \int_0^z \frac{dz'}{\sqrt{1 - \Omega_0^\mathrm{(tot)} + 
\Omega_0(1+z')^{3(w+1)}  }} \nonumber\\
&&\nonumber\\
&=& \int_x^1 dx' \left( x' \right)^{-2} \left[ 1-\Omega_0^\mathrm{(tot)} 
+\Omega_0 (x')^{-3(w_1+1)} \right]^{-1/2} \,,
\end{eqnarray}
which is of the Chebyshev form~(\ref{integral}) with $ p=-2$, $r=-3(w+1)$, 
$ q=-1/2$. 
Imposing that $w\in \mathbb{Q}$, it is 
\be
\frac{p+1}{r} = \frac{1}{3(w+1)} \,, \quad \frac{p+1}{r}+q = 
-\frac{(3w+1)}{6(w+1)}
\ee
and
$ (p+1)/r  = n \in  \mathbb{Z}$ implies 
\begin{eqnarray}
&&w_n=\frac{1}{3n}-1 = -1, \, ... \, -\frac{10}{9}, -\frac{ 7}{6}, - 
\frac{4}{3} , -\frac{2}{3} , -\frac{5}{6}, - \frac{8}{9}, \, ... \, , -1 
\,.\nonumber\\
&&
\end{eqnarray}
The second condition $ \frac{p+1}{r}+q=m\in \mathbb{Z}$ yields
\begin{eqnarray}
w_m &=& - \frac{(6m+1)}{3(2m+1)}= 
 -1, \, ... \, , -\frac{1}{3}, -\frac{7}{9}, -\frac{5}{3}, 
\frac{13}{9}, \, ... \, , -1 \nonumber\\
&&
\end{eqnarray}
or $ n=0$ and $w=-1/3 $, which  corresponds to the Milne universe 
(Minkowski space with a hyperbolic foliation).

The situation of a single fluid plus cosmological constant $\Lambda$ is 
obtained with the replacement $ 1-\Omega_0^\mathrm{(tot)} \to 1- 
\Omega_0^\mathrm{(tot)} +\Omega_{\Lambda 0}$.

\subsection{Two fluids}

Suppose that the FLRW universe is sourced by two fluids, then $D_L(z)$ 
depends on 
\begin{eqnarray}
&& \int_0^z \frac{ dz'}{ \sqrt{
1-\Omega_0^\mathrm{(tot)} 
+\Omega_0^{(1)} (1+z')^{3(w_1+1)} 
+\Omega_0^{(2)} (1+z')^{3(w_2+1)} } } \nonumber\\
&& \equiv I_2 \,.
\end{eqnarray}
For $K=0$ (or $\Omega_0^\mathrm{(tot)}=1$), corresponding to the 
luminosity distance~(\ref{DLK=0}), use the variable $y \equiv 
1+z$ to obtain
\begin{eqnarray}
I_2 &=& \int_1^y \frac{dy'}{\sqrt{ \Omega_0^{(1)} y^{ 3(w_1+1)} 
+\Omega_0^{(2)} \, y^{3(w_2+1)} }} \nonumber\\
&&\nonumber\\
&=& \int_1^y dy' \frac{ (y')^{\frac{-3(w_1+1)}{2} }  }{\sqrt{ 
\Omega_0^{(1)} 
+\Omega_0^{(2)} (y')^{ 3(w_2-w_1)} } } \,,
\end{eqnarray}
which is of the form~(\ref{integral}) appearing in the Chebyshev theorem 
with $p=-3(w_1+1)/2$, ~$q=-1/2$, and $r=3(w_2-w_1)$. Then
\be
\frac{p+1}{r} = \frac{-(3w_1+1)}{6(w_2-w_1)} \,, \quad \quad \frac{p+1}{r} 
+q=\frac{-(3w_2+1)}{6(w_2-w_1)} \,. 
\ee
The first condition for ``simple'' integrability $(p+1)/r=n \in 
\mathbb{Z}$ gives
\begin{eqnarray}
w_2= \frac{3(2n-1)w_1-1}{6n} \quad \mbox{or} \quad 
w_1=-\frac{1}{3} 
\,, \quad \mbox{any} \; w_2 \neq - \frac{1}{3} \,.\nonumber\\
&&
\end{eqnarray}
If the first fluid is a dust ($w_1=0$), we obtain the pairs 
\be
\begin{array}{rcll}
\left( w_1, w_2 \right) &=& \left( 0, 0 \right)& \mbox{(single 
dust fluid)},\\
&&&\\
&  ... &  &\\
&&&\\
\left( w_1, w_2 \right) &=& \left( 0, \pm \frac{1}{6}  \right) ,& \\
&&&\\
\left( w_1, w_2 \right) &=& \left( 0,  \pm \frac{1}{12} \right) ,&\\
&&&\\
\left( w_1, w_2 \right) &=& \left( 0, -\frac{1}{18} \right) ,&\\
&&&\\
\left( w_1, w_2 \right) &=& \left( 0, \frac{1}{9} \right) ,&\\
&&&\\
&  ... & &\\
&&&\\
\left( w_1, w_2 \right) &=& \left( 0 , 0 \right) & \mbox{(again, a 
single dust)}.
\end{array}
\ee

If instead the first fluid is radiation, $w_1=1/3$, we have the pairs
\be
\begin{array}{rcll}
\left( w_1, w_2 \right) &=& \left( \frac{1}{3}, \frac{1}{3} \right)& 
\mbox{(single radiation fluid)},\\
&&&\\
&  ... &  &\\
&&&\\
\left( w_1, w_2 \right) &=& \left( \frac{1}{3}, 0 \right) & 
\mbox{(radiation plus dust)}, \\
&&&\\
\left( w_1, w_2 \right) &=& \left( \frac{1}{3},  \frac{2}{3} \right) ,&\\
&&&\\
\left( w_1, w_2 \right) &=& \left( \frac{1}{3}, \frac{1}{6} \right) ,&\\
&&&\\
\left( w_1, w_2 \right) &=& \left( \frac{1}{3}, \frac{1}{2} \right) , &   
\\
&&&\\
\left( w_1, w_2 \right) &=& \left( \frac{1}{3}, \frac{2}{9} \right) , &\\
&&&\\
\left( w_1, w_2 \right) &=& \left( \frac{1}{3}, \frac{4}{9} \right) ,&\\
&&&\\
&  ... & &\\
&&&\\
\left( w_1, w_2 \right) &=& \left( \frac{1}{3} , \frac{1}{3} \right) & 
\mbox{(again, a single radiation fluid)}.
\end{array} 
\ee

If the first fluid is stiff, $w_1=1$, the pairs giving Chebyshev 
integrability are 
\be
\begin{array}{rcll}
\left( w_1, w_2 \right) &=& \left( 1, 1 \right)& 
\mbox{(single stiff fluid)},\\
&&&\\
&  ... &  &\\
&&&\\
\left( w_1, w_2 \right) &=& \left( 1, \frac{1}{3} \right) & 
\mbox{(stiff fluid plus radiation)}, \\
&&&\\
\left( w_1, w_2 \right) &=& \left( 1,  \frac{5}{3} \right) ,&\\
&&&\\
\left( w_1, w_2 \right) &=& \left( 1, \frac{2}{3} \right) ,&\\
&&&\\
\left( w_1, w_2 \right) &=& \left( 1, \frac{4}{3} \right) , &   
\\
&&&\\
\left( w_1, w_2 \right) &=& \left( 1, \frac{7}{9} \right) , &\\
&&&\\
&  ... & &\\
&&&\\
\left( w_1, w_2 \right) &=& \left( 1 , 1 \right) & 
\mbox{(again, a single stiff fluid)}.
\end{array} 
\ee

If the first (effective) fluid is the cosmological constant, $w_1=-1$, we 
have instead the pairs 
\be
\begin{array}{rcll}
\left( w_1, w_2 \right) &=& \left( -1, 1 \right)& 
( \Lambda \; \mbox{and no matter}),\\
&&&\\
&  ... &  &\\
&&&\\
\left( w_1, w_2 \right) &=& \left( -1, -\frac{2}{3} \right) & 
\mbox{(stiff fluid plus radiation)}, \\
&&&\\
\left( w_1, w_2 \right) &=& \left( -1, -\frac{4}{3} \right) ,&\\
&&&\\
\left( w_1, w_2 \right) &=& \left( -1, -\frac{5}{6} \right) ,&\\
&&&\\
\left( w_1, w_2 \right) &=& \left( -1, -\frac{7}{6} \right) , &   
\\
&&&\\
&  ... & &\\
&&&\\
\left( w_1, w_2 \right) &=& \left( -1 , -1 \right) & 
\mbox{(again}\:  \Lambda \: \mbox{and no matter}).
\end{array} 
\ee

The second condition for integrability \`{a} la Chebyshev
\be
\frac{p+1}{r} +q=\frac{-(3w_2+1)}{6(w_2-w_1)} =m \in \mathbb{Z}
\ee
gives 
\be
w_2=\frac{6mw_1-1}{3(2m+1)} \quad\mbox{or}\quad w_2=-\frac{1}{3}\,, 
\quad\mbox{any}\: w_1\neq -\frac{1}{3} \,.
\ee

In particular, if the first fluid is a dust, $w_1=0$, we have 
$w_2=-\left([ 3(2m+1)\right]^{-1}$ and the pairs

\be
\begin{array}{rcll}
\left( w_1, w_2 \right) &=& \left( 0, 0 \right)& \mbox{(single 
dust fluid)},\\
&&&\\
&  ... &  &\\
&&&\\
\left( w_1, w_2 \right) &=& \left( 0, \pm \frac{1}{3}  \right) ,& \\
&&&\\
\left( w_1, w_2 \right) &=& \left( 0,  \pm \frac{1}{9} \right) ,&\\
&&&\\
\left( w_1, w_2 \right) &=& \left( 0, \pm \frac{1}{15} \right) ,&\\
&&&\\
&  ... & &\\
&&&\\
\left( w_1, w_2 \right) &=& \left( 0 , 0 \right) & \mbox{(again, a 
single dust)}.
\end{array}
\ee

If the first fluid is radiation, $w_1=1/3$, then 
$w_2=\frac{2m-1}{3(2m+1)}$, giving the pairs

\be
\begin{array}{rcll}
\left( w_1, w_2 \right) &=& \left( \frac{1}{3}, \frac{1}{3} \right)& 
\mbox{(single radiation fluid)},\\
&&&\\
&  ... &  &\\
&&&\\
\left( w_1, w_2 \right) &=& \left( \frac{1}{3}, \pm 1 \right) & 
(\mbox{radiation plus dust or}\, \Lambda)  , \\
&&&\\
\left( w_1, w_2 \right) &=& \left( \frac{1}{3},  -\frac{1}{3} \right) ,&\\
&&&\\
\left( w_1, w_2 \right) &=& \left( \frac{1}{3}, \frac{1}{9} \right) ,&\\
&&&\\
\left( w_1, w_2 \right) &=& \left( \frac{1}{3}, \frac{1}{5} \right) , &   
\\
&&&\\
\left( w_1, w_2 \right) &=& \left( \frac{1}{3}, \frac{5}{9} \right) , &\\
&&&\\
\left( w_1, w_2 \right) &=& \left( \frac{1}{3}, \frac{5}{21} \right) ,&\\
&&&\\
&  ... & &\\
&&&\\
\left( w_1, w_2 \right) &=& \left( \frac{1}{3} , \frac{1}{3} \right) & 
\mbox{(again, a single radiation fluid)}.
\end{array} 
\ee

If the first fluid is stiff with $w_1=1$, then $ w_2= \frac{6m-1}{3(2m+1)} 
$, generating the pairs

\be
\begin{array}{rcll}
\left( w_1, w_2 \right) &=& \left( 1, 1 \right)& 
\mbox{(single stiff fluid)},\\
&&&\\
&  ... &  &\\
&&&\\
\left( w_1, w_2 \right) &=& \left( 1, -\frac{1}{3} \right) & 
\mbox{(stiff fluid plus radiation)}, \\
&&&\\
\left( w_1, w_2 \right) &=& \left( 1,  \frac{1}{9} \right) ,&\\
&&&\\
\left( w_1, w_2 \right) &=& \left( 1, \frac{7}{13} \right) ,&\\
&&&\\
\left( w_1, w_2 \right) &=& \left( 1, \frac{11}{15} \right) , &   
\\
&&&\\
\left( w_1, w_2 \right) &=& \left( 1, \frac{13}{9} \right) , &\\
&&&\\
&  ... & &\\
&&&\\
\left( w_1, w_2 \right) &=& \left( 1 , 1 \right) & 
\mbox{(again, a single stiff fluid)}.
\end{array} 
\ee

As an example, consider the case of a spatially flat FLRW 
universe filled with radiation and dust, $w_1=1/3$ and 
$w_2=0$ appearing in the list~(\ref{list2}), in which case
\begin{eqnarray}
I_2 &=& \int_1^y dy' \, \frac{ (y')^{-2}}{ \sqrt{ \Omega_0^{(1)} 
+\Omega_0^{(2)} /y'}}\nonumber\\
&&\nonumber\\
& =& -\frac{2}{\Omega_0^{(2)} } \sqrt{ \Omega_0^{(1)} 
+\frac{\Omega_0^{(2)}}{y'} } \, \Bigg|_1^y \nonumber\\
&&\nonumber\\
&=&  \frac{2}{\Omega_0^{(2)} } \left[ 
\sqrt{ \Omega_0^{(1)} +\Omega_0^{(2)} }    -
\sqrt{ \Omega_0^{(1)} + \frac{\Omega_0^{(2)}}{y} } \,  \right]
\end{eqnarray}
and the luminosity distance versus redshift relation is
\begin{eqnarray}
D_L(z) &=& \frac{2 H_0^{-1} \left(1+z\right)}{\Omega_0^\mathrm{(dust)} 
}\nonumber\\
&&\nonumber\\
&\, & \times \left[ \sqrt{ \Omega_0^\mathrm{(rad)} 
+\Omega_0^\mathrm{(dust)} }-
\sqrt{ \Omega_0^\mathrm{(rad)} + \frac{\Omega_0^\mathrm{(dust)}}{1+z} } 
\,\, \right] \,.\nonumber\\
&&
\end{eqnarray}

It is unfortunate that the $\Lambda$CDM model corresponding to $\Lambda$ 
and dust is not integrable \`a la Chebyshev. Usually, the luminosity 
distance $D_L(z)$ is expanded for small $z$ to compare it with type Ia 
supernovae data, However, standard candles at redshifts $z\sim 1$ are 
present in current catalogues and the small $z$ expansion fails for those 
objects, hence the search for new parametrizations valid at high redshifts 
\cite{Chevallier:2000qy,Linder:2002et,Cattoen:2007sk}.

\section{Conclusions}
\label{sec:5}

It is of interest to know when the lookback time $t_L$, the age $t_0$ of 
the universe, and the luminosity distance versus redshift $D_L(z)$ can be 
computed analytically in FLRW cosmology.  These quantities contain 
integrals expressed by hypergeometric series, which truncate to a finite 
number of terms under certain conditions expressed by the Chebyshev 
theorem of integration \cite{Chebyshev,MarchisottoZakeri}. We have 
classified the situations in which the Chebyshev theorem holds for a FLRW 
universe containing real or effective fluids (including curvature and the 
cosmological constant). The Chebyshev theorem is not useful for situations 
with more than three fluids or effective fluids. Moreover, when the 
universe is dominated by a single fluid for most of its history, one can 
approximate the age of the universe with the duration of the epoch 
dominated by that fluid (for example, in a universe containing only dust 
and radiation, with $\Lambda =0$, neglecting the duration of the 
radiation-dominated age only introduces a small error in the age  computed 
using only dust).

In cosmography, the luminosity distance versus redshift relation has been 
instrumental in detecting the acceleration of the cosmic expansion with 
type Ia supernovae 
\cite{SupernovaSearchTeam:1998fmf,SupernovaCosmologyProject:1998vns,Filippenko:1998tv,Riess:1999ti,Riess:2000yp,SupernovaSearchTeam:2001qse,SupernovaSearchTeam:2003cyd,SupernovaCosmologyProject:2003dcn,Barris:2003dq,SupernovaSearchTeam:2004lze,Riess:2019qba} 
and is one of the most important observational relations. Building 
observational plots of $D_L$ versus $z$ relies on expanding the relation 
$D_L(z)$ to second order around the present time and measuring the present 
values $H_0 \equiv \dot{a}/a \Big|_0$ of the Hubble function and $q_0 
\equiv - \ddot{a} a / \dot{a}^2 \Big|_0 $ of the deceleration parameter 
(the third and fourth order terms in the series or, equivalently, the jerk 
and the snap are subject to much larger uncertainties). When distant 
objects at redshift $z \sim 1 $ are included in the samples, the expansion 
breaks down and one has to resort to alternative parametrizations, for 
example the Chevallier-Polarski-Linder (CPL) 
\cite{Chevallier:2000qy,Linder:2002et} or the Cattoen-Visser 
\cite{Cattoen:2007sk} parametrizations. Being able to compute exactly 
$D_L(z)$ is complementary to the cosmographic and numerical approaches. 
Unfortunately, among the infinitely many cases in which integration \`a la 
Chebyshev is possible, only a few correspond to physically realistic 
situations or even (real or effective) realistic fluids. Nevertheless, one 
wants to know when simple analytical expressions of $t_0$ and $D_L(z)$ 
exist. Even when they do not describe realistic epochs of the history of 
the universe, these situations can be used as toy models for theoretical 
purposes or for testing parametrizations in cosmography or numerical 
evaluations of $t_L$, $t_0$, and $D_L(z)$.

\begin{acknowledgements}

This work is supported by the Natural Sciences \& Engineering Research 
Council of Canada (grant no. 2016-03803 to V.F.) and by a Bishop's 
University Graduate Entrance Scholarship (S.J.).

\end{acknowledgements}


\appendix
\section{Lookback time and age for $K=0$, $ \Lambda\neq 0$, and a single 
fluid}
\label{sec:appendix}

The lookback time~(\ref{strakaz}) integrates to 
\begin{eqnarray}
t_L &=& \frac{2H_0^{-1} }{ 3(w+1)\sqrt{\Omega_{\Lambda 0}} } \coth^{-1} 
\Bigg( \frac{ 
\sqrt{ \Omega_{\Lambda 0} +\Omega_0 \, x^{-3(w+1)} } }{
\sqrt{\Omega_{\Lambda 0} }} \Bigg) \Bigg|_{x_e}^1 \,.\nonumber\\
&&
\end{eqnarray}
Using the identity
\be
\coth^{-1} z=\frac{1}{2} \, \ln \left( \frac{z+1}{z-1} \right) 
\ee
for $|z|>1$, we have 
\begin{eqnarray}
t_L &=& \frac{2H_0^{-1} }{ 3(w+1)\sqrt{\Omega_{\Lambda 0} } }
\, \frac{1}{2} \, \ln \Bigg(
\frac{  
\frac{\sqrt{ \Omega_{\Lambda 0} + \Omega_0 \, x^{-3(w+1)} } }{
\sqrt{\Omega_{\Lambda 0} }} +1}{
\frac{\sqrt{ \Omega_{\Lambda 0}+ \Omega_0 \, x^{-3(w+1)} }}{
\sqrt{\Omega_{\Lambda 0}}} -1} \Bigg) \Bigg|_{x_e}^1 \nonumber\\
&&\nonumber\\  
&=& \frac{H_0^{-1} }{3(w+1)\sqrt{\Omega_{\Lambda 0} }  } \left[
\ln \Bigg( \frac{ 1+ \sqrt{ \Omega_{\Lambda 0}} }{
1-\sqrt{ \Omega_{\Lambda 0} }  } \Bigg) \right.\nonumber\\
&&\nonumber\\
&\, & \left. -\ln \Bigg( \frac{
\sqrt{ \Omega_{\Lambda 0} \, x^{3(w+1)} + \Omega_0 } +
\sqrt{ \Omega_{\Lambda 0} \, x^{3(w+1)}  } }{
 \sqrt{ \Omega_{\Lambda 0}\,  x^{3(w+1)} + \Omega_0 } - 
\sqrt{ \Omega_{\Lambda 0}\,  x^{3(w+1)}  } } \Bigg) \right]\,.
\end{eqnarray}  
In the limit $x_e\to 0$ one finds
\begin{eqnarray}
t_L & \to & t_0 = \frac{H_0^{-1} }{3(w+1)\sqrt{\Omega_{\Lambda 0} } } \, 
\ln \Bigg( 
\frac{ 1+ \sqrt{ \Omega_{\Lambda 0}} }{  1-\sqrt{ \Omega_{\Lambda 0} }} 
\Bigg) \nonumber\\ 
&&\nonumber\\
&= & \frac{2H_0^{-1}}{3(w+1)\sqrt{\Omega_{\Lambda 0}}} \, \frac{1}{2} \, 
\ln \Bigg[ 
\frac{ \left( 1+ \sqrt{ \Omega_{\Lambda 0}} \right)^2 }{  1- 
\Omega_{\Lambda 0} } \Bigg] \nonumber\\ 
&& \nonumber\\
&=& \frac{2H_0^{-1}}{3(w+1)\sqrt{\Omega_{\Lambda 0}}} \, \ln \Bigg( 
\frac{ 1+ \sqrt{ \Omega_{\Lambda 0}} }{ \sqrt{ 1- 
\Omega_{\Lambda 0} }} \Bigg) \,.
\end{eqnarray}


\begin{thebibliography}{}

\bibitem{Chebyshev} P.~L. Chebyshev, ``Sur l’integration des 
diff\'erentielles irrationnelles'', J. Mathematiques (series
1) {\bf 18}, 87–111 (1853).

\bibitem{MarchisottoZakeri} E.~A. Marchisotto, G.-A. Zakeri, ``An 
invitation to integration in finite terms'', Coll. Math. J.  {\bf 25}, 
295–308 (1994).

\bibitem{Jacobs1968} K.~C. Jacobs, ``Spatially homogeneous and Euclidean 
cosmological models with shear'', Astrophys. J. {\bf 153}, 661-678 (1968).

\bibitem{Vajik69} J.~P. Vajk, ``Exact Robertson‐Walker Cosmological 
Solutions Containing Relativistic Fluids'', J. Math. Phys. {\bf 10}, 
1145-1151 (1969).

\bibitem{McIntosh1972} C.~B.~G. McIntosh, ``I.~Robertson-Walker metric'',  
Austral. J. Phys. {\bf 25}, 75-82 (1972).

\bibitem{McIntoshFoyster1972} C.~B.~G. McIntosh, J.~M. Foyster, 
``Cosmological models with two fluids~II. Conformal and conformally flat 
metrics'', Austral. J. Phys. {\bf 25}, 83-89 (1972).
         
\bibitem{Chen:2014fqa} S.~Chen, G.~W.~Gibbons, Y.~Li, Y.~Yang, 
``Friedmann's Equations in All Dimensions and Chebyshev's Theorem'', JCAP 
\textbf{12}, 035 (2014) doi:10.1088/1475-7516/2014/12/035 [arXiv:1409.3352 
[astro-ph.CO]].

\bibitem{Faraoni:2021opj}
V.~Faraoni, S.~Jose and S.~Dussault,
``Multi-fluid cosmology in Einstein gravity: analytical solutions,''
Gen. Rel. Grav. \textbf{53}, no.12, 109 (2021)
doi:10.1007/s10714-021-02879-z
[arXiv:2107.12488 [gr-qc]].

\bibitem{Waldbook} R.~M. Wald, {\em General Relativity} (Chicago 
University Press, Chicago, 1984).

\bibitem{EllisMaartensMacCallum} G.~F.~R. Ellis, R. Maartens, 
and M.~A.~H. MacCallum, {\em Relativistic cosmology} (Cambridge 
University Press, Cambridge, 2012).

\bibitem{KolbTurner} E. W. Kolb and M. S. Turner, {\em The Early Universe} 
(Addison-Wesley, Redwood City, CA, 1990).

\bibitem{Slava} V. Mukhanov, {\it Physical Foundations of Cosmology} 
(Cambridge University Press, Cambridge, 2005), 
doi:10.1017/CBO9780511790553. 

\bibitem{Ellis:1998ct}
G.~F.~R.~Ellis and H.~van Elst,
``Cosmological models: Cargese lectures 1998,''
NATO Sci. Ser. C \textbf{541}, 1-116 (1999)
doi:10.1007/978-94-011-4455-1\_1
[arXiv:gr-qc/9812046 [gr-qc]].

\bibitem{Mattig58} W. Mattig, ``\"Uber der zusammenhang zwischen 
rotverschiebung und scheinrare helligkeit'', Astron. Nachr. \textbf{284}, 
109 (1958).

\bibitem{SupernovaSearchTeam:1998fmf}
A.~G.~Riess \textit{et al.} [Supernova Search Team],
``Observational evidence from supernovae for an accelerating universe and 
a cosmological constant,''
Astron. J. \textbf{116}, 1009-1038 (1998)
doi:10.1086/300499
[arXiv:astro-ph/9805201 [astro-ph]].

\bibitem{SupernovaCosmologyProject:1998vns}
S.~Perlmutter \textit{et al.} [Supernova Cosmology Project],
``Measurements of $\Omega$ and $\Lambda$ from 42 high redshift 
supernovae,''
Astrophys. J. \textbf{517}, 565-586 (1999)
doi:10.1086/307221
[arXiv:astro-ph/9812133 [astro-ph]].

\bibitem{Filippenko:1998tv}
A.~V.~Filippenko and A.~G.~Riess,
``Results from the high Z supernova search team,''
Phys. Rept. \textbf{307}, 31-44 (1998)
doi:10.1016/S0370-1573(98)00052-0
[arXiv:astro-ph/9807008 [astro-ph]].

\bibitem{Riess:1999ti}
A.~G.~Riess, A.~V.~Filippenko, W.~Li and B.~P.~Schmidt,
``An indication of evolution of type ia supernovae from their 
risetimes,''
Astron. J. \textbf{118}, 2668-2674 (1999)
doi:10.1086/301144
[arXiv:astro-ph/9907038 [astro-ph]].

\bibitem{Riess:2000yp}
A.~G.~Riess,
``The case for an accelerating universe from supernovae,''
Publ. Astron. Soc. Pac. \textbf{112}, 1284 (2000)
doi:10.1086/316624
[arXiv:astro-ph/0005229 [astro-ph]].

\bibitem{SupernovaSearchTeam:2001qse}
A.~G.~Riess \textit{et al.} [Supernova Search Team],
``The farthest known supernova: support for an accelerating universe and 
a glimpse of the epoch of deceleration,''
Astrophys. J. \textbf{560}, 49-71 (2001)
doi:10.1086/322348
[arXiv:astro-ph/0104455 [astro-ph]].

\bibitem{SupernovaSearchTeam:2003cyd}
J.~L.~Tonry \textit{et al.} [Supernova Search Team],
``Cosmological results from high-z supernovae,''
Astrophys. J. \textbf{594}, 1-24 (2003)
doi:10.1086/376865
[arXiv:astro-ph/0305008 [astro-ph]].

\bibitem{SupernovaCosmologyProject:2003dcn}
R.~A.~Knop \textit{et al.} [Supernova Cosmology Project],
``New constraints on $\Omega_M, \Omega_{\Lambda}$, and $w$ from an 
independent set of eleven high-redshift supernovae observed with HST,''
Astrophys. J. \textbf{598}, 102 (2003)
doi:10.1086/378560
[arXiv:astro-ph/0309368 [astro-ph]].

\bibitem{Barris:2003dq}
B.~J.~Barris, J.~L.~Tonry, S.~Blondin, P.~Challis, R.~Chornock, 
A.~Clocchiatti, A.~V.~Filippenko, P.~Garnavich, S.~T.~Holland and S.~Jha, 
\textit{et al.}
``23 High redshift supernovae from the IFA Deep Survey: Doubling the SN 
sample at z \ensuremath{>} 0.7,''
Astrophys. J. \textbf{602}, 571-594 (2004)
doi:10.1086/381122
[arXiv:astro-ph/0310843 [astro-ph]].

\bibitem{SupernovaSearchTeam:2004lze}
A.~G.~Riess \textit{et al.} [Supernova Search Team],
``Type Ia supernova discoveries at z \ensuremath{>} 1 from the Hubble 
Space Telescope: Evidence for past deceleration and constraints on dark 
energy evolution,''
Astrophys. J. \textbf{607}, 665-687 (2004)
doi:10.1086/383612
[arXiv:astro-ph/0402512 [astro-ph]].

\bibitem{Riess:2019qba}
A.~G.~Riess,
``The Expansion of the Universe is Faster than Expected,''
Nature Rev. Phys. \textbf{2}, no.1, 10-12 (2019)
doi:10.1038/s42254-019-0137-0
[arXiv:2001.03624 [astro-ph.CO]].

\bibitem{Bassett:2003vu}
B.~A.~Bassett and M.~Kunz,
``Cosmic distance-duality as a probe of exotic physics and 
acceleration,''
Phys. Rev. D \textbf{69}, 101305 (2004)
doi:10.1103/PhysRevD.69.101305
[arXiv:astro-ph/0312443 [astro-ph]].

\bibitem{Carroll} S. Carroll, {\em An Introduction to General Relativity} 
(Addison-Wesley, San Francisco, 2004).

\bibitem{Chevallier:2000qy}
M.~Chevallier and D.~Polarski,
``Accelerating universes with scaling dark matter,''
Int. J. Mod. Phys. D \textbf{10}, 213-224 (2001)
doi:10.1142/S0218271801000822
[arXiv:gr-qc/0009008 [gr-qc]].

\bibitem{Linder:2002et}
E.~V.~Linder,
``Exploring the expansion history of the universe,''
Phys. Rev. Lett. \textbf{90}, 091301 (2003)
doi:10.1103/PhysRevLett.90.091301
[arXiv:astro-ph/0208512 [astro-ph]].

\bibitem{Cattoen:2007sk}
C.~Cattoen and M.~Visser,
``The Hubble series: Convergence properties and redshift variables,''
Class. Quant. Grav. \textbf{24}, 5985-5998 (2007)
doi:10.1088/0264-9381/24/23/018
[arXiv:0710.1887 [gr-qc]].

\end{thebibliography}


\end{document}